# New Index for Quantifying an Individual's Scientific Research Output

M. Abdel-Aty

Scientific Publishing Center, 32038 University of Bahrain, Kingdom of Bahrain
Mathematics Department, Faculty of Science, Sohag University, 82524 Sohag, Egypt

**Abstract:** Classifying researchers according to the quality of their published work rather than the quantity is a curtail issue. We attempt to introduce a new formula of the percentage range to be used for evaluating qualitatively the researchers' production. The suggested equation depends on the number of the single-author published papers and their citations to be added as a new factor to the known *h*-index. These factors give an advantage and make a clear evidence of innovative authors and reduce the known *h*-index for authors who are gaining citations by adding their names to multi-author papers. It is shown that various dimensions of ethical integrity and originality will be effective in this new index. An important scenario arising from the analysis is shown in terms of examples. It refers to larger differences between the *h- and the new*index which comes from the whole work and the one comes from the single-author papers only, is shown.

## Introduction

We would like to start this paper by the following question: are we judging the researchers production by the correct standards?. Although qualitative method such as *h*-index being used and grown increasingly by different institutions and different dimensions, confusion exists over how different authors with different contributions hav ethe same *h*-index and how quality can be assessed when authors with multi-author papers compared to authors with single-author papers [Bakkalbasi, Bauer, Glover, Wang, (2006)]. These topics will be addressed in this paper.

In 2005, Hirsch(2005)has --- proposed an indicator which is called the *h*-index, defined as the number of papers with citation number higher or equal to *h*, as a useful index to characterize the scientific output of a researcher i.e. a scientist has index *h* if *h* of his/her *Np* papers have at least *h* citations each, and the other (*Np-h*) papers have no more than *h* citations each. This index is used to characterize the scientific output of a researcher in a very good way. The way of calculation of the *h*-index includes the total number of papers published over a certain period of years and the number of citations for each paper. Now the Institute of Scientific Information(ISI) Web of Knowledge indexes more than 11,000 science and social science journals [Thomson Reuters. 2010.] and uses the impact factor to report on ranking the journals. Elsevier as an international publishers established the independent and international Scopus [Scopus Info. Elsevier. Retrieved 2013-01-29] content selection and advisory board to avoid any conflict of interest in the choice of the journals to be considered for the inclusion in the database and to maintain an open and transparent content coverage policy. Scopus covers nearly 20,500 titles from over 5,000 international publishers, of which 19,500 are peer-reviewed journals in the scientific, technical, medical, and social sciences [Scopus Info. Elsevier. Retrieved 2013-01-29, Elsevier introduces SciVerse,(2010),see also Falagas, Pitsouni, Malietzis, Pappas(2007)].

Egghe(2006)suggested what is called *g*-index. This index is used to quantifying scientific productivity based on publication records and is calculated through the distribution of citations received by a given researcher's publications. Given a set of articles, ranked in decreasing order of the number of citations that they received, the *g*-index is the (unique) largest number such that the top *g* articles received (together) at least $g^2$ citations. In agreement with the *h*-index, the *g*-index is a number close to the *h*-index for the same authors.

An application of the modified version of the *h*-index and impact factor has been used in 2012 [Abdel-Aty, 2012], to estimate of the impact of Journals published in the Arabic Language as well as Arabic scientist's cumulative research contributions. In order to achieve the strongest indicators for any journal or researcher, a model system has been developed in order to determine numerically an index. Consider an individual researcher: it has been suggested to use a modified version of the Hirschindex: ----

$$AsF = \left(\frac{h}{h+1}\right) \times 100.$$

From this equation it is shown that AsF is an increasing number according to the increment of the *h*-index but does not exceed 100. In particular, assume that the researcher has *h*-index of 1, this corresponds to *AsF=50%* and *h*-index *5* corresponds to *AsF=83.3%* and so on. When the *h*-index is rising then the difference with respect to the *AsF* is getting smaller.

The issue we have in mind has to do with the researchers quality and index meaning, taking into account the dependence on the relevant magnitudes such as the number of single-author papers, the number of citations and the *h*-index [Hirsch,2005]. This is most conveniently accomplishedin a mathematical formalism in terms of one equation. Related treatments based on both *h*-index and number of citations, discussing authors ranks, without the spatial dependence, have been presented in the literature [Abdel-Aty, 2012].What we have studied and present below is essentially the most general form of the complete indexing equation.

Now, one may ask the following question: is there any difference of the *h*-index for two authors, each one of them has published the same number of papers and those papers have received the same number of citations, but one of them has published most of his papers as single-author papers and the second published all his papers as multi-author papers and even his contribution is very limited in all these papers?. The answer according to the present index is: yes, they have the same *h*-index. This means that we need to add some factors to the current index to avoid such misleading result.

There is a debate in the literature about whether the concepts of h-index as a quality factor is used toassess qualitative output should be roughly the same as, parallel to, or quite different from the results obtained from the citations [Hollway, (2007)a,b]. In fact the quality of the published papers comes from the work itself and how the author benefit from this work should depends on his contribution but need to be reformulated and assessed quite differently within the domain of different factors. We suggest new equation to be clear, thoughtful and reflexive about the impact of different variables which cannot be avoided when we speak about the research quality. At this stage we introduce a new index. This index takes into account the total number of published papers, the number of citations, the *h*-index for both single-author papers and multi-author papers, in the following form

$$A = \sum_{i=1}^{6} \mu(i),$$

$$\mu(1) = \frac{N_1}{N_1 + 1} \times \Omega_1,$$

$$\mu(2) = \Omega_2 \times \frac{max\left(0, \frac{N_2 - 100}{200}\right)}{\left[1 + max\left(0, \frac{N_2 - 100}{200}\right)\right]},$$

$$\mu(3,4) = \frac{N_{3,4}}{N_{3,4} + 1} \times \Omega_{3,4},$$

$$\mu(5) = \Omega_5 \times \frac{max\left(0, \frac{N_5 - 70}{50}\right)}{\left[1 + max\left(0, \frac{N_5 - 70}{50}\right)\right]},$$

$$\mu(6) = \frac{N_6}{N_6 + 1} \times \Omega_6.$$

where, $N_1$, is the total number of published papers by the author including the multi-author papers,

$N_2$ is the modified total number of the citations which has been received for all published papers by the author, including the multi-author papers,

$N_3$ is the h-index of the author for all his work,

$N_4$ is the number of single-author papers published in the ISI journals by the author,

$N_5$ is the modified number of citations received for the single-author papers only,

$N_6$ is the author's h-index according to the single-author papers only. According the above mentioned weights, $\Omega_1 = 20$, $\Omega_2 = 10$, $\Omega_{3,4} = 14$, $\Omega_5 = 12$ and $\Omega_6 = 30$.

From now on we will call the above equation just the *A*-index. The above equation offers the chance to more factors to play their necessary roles in the final results of quantifying the individual's scientific research output. Also, the weight of each factor can be changed according to the purpose of the evaluation.

Example: Two authors each of them have published 20 papers in ISI journals, 11 of them have been cited 11 or more and the other information are

|  | No. of Published papers | No of Citations | h-index | No. of Single-author papers | Citations for single author-papers | h-index of the single-author papers |
|---|---|---|---|---|---|---|
| Author 1 | 20 | 300 | 11 | 0 | 0 | 0 |
| Author 2 | 20 | 300 | 11 | 15 | 300 | 8 |

from the above table, we see that the two authors have the same h-index, even the first one did not publish any single-author paper and the second one has published most of his work independently (single-author papers) and her/his work is well recognized, where the single-author papers have received 300 citations and finally according to her/his independent work she/he has *h*-index of 8 and the first one has *h*-index of zero.

According to our equation, the index of the first one is 42.06 out of 100 while the second author obtained 97.05 out of 100. Which means that the number of single-author papers and their citations play an important role in the evaluation of the quality of the researchers outcome.

It has been shown recently that Web of Science, as well as Scopus and Google Scholar produced results for citations, quantitatively and qualitatively, different for articles published in some different journals [Kulkarni, Aziz, Shams, Busse, (2009)]. They mentioned to the fact that different citation data bases, including Web of Science, Scopus and Google Scholar use unique methods to record and count citations and have shown differences in citation counts for some different fields [Harzing, van der Wal, (2008)]. In this regard, our equation gives a reasonable solution and an acceptable index for all cases and adds some points for each factor.

Example 2: In the following example we ---show the effect of each variable when all other variables are kept constant.

|  | No. of Published papers | No of Citations | h-index | No. of Single-author papers | Citations for single author-papers | h-index of the single-author papers |
|---|---|---|---|---|---|---|
| Author 1 | 20 | 200 | 11 | 5 | 100 | 3 |
| Author 2 | 20 | 400 | 11 | 5 | 100 | 3 |
| Author 3 | 20 | 800 | 11 | 5 | 100 | 3 |
| Author 4 | 20 | 3000 | 11 | 5 | 100 | 3 |

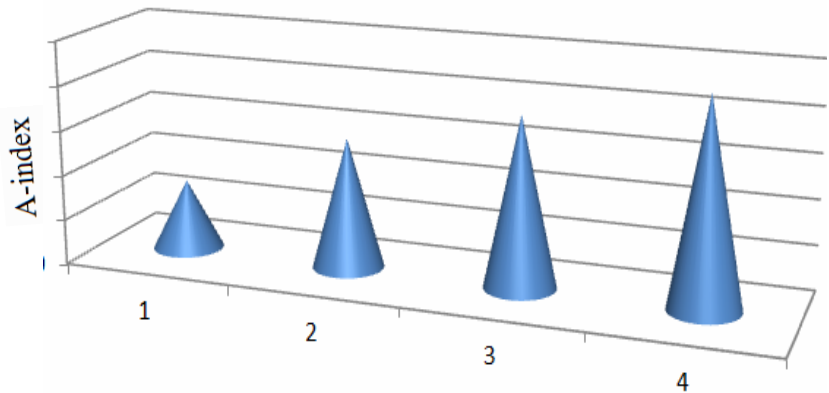

Fig. 1, plot of A-index for different values of the number of total citations.

| Authors | A1 | A2 | A3 | A4 |
|---|---|---|---|---|
| Total No. of Citations | 200 | 400 | 800 | 3000 |
| A-index | 73.131 | 75.797 | 77.575 | 79.152 |

In this example we fix all parameters and change only the total number of citations, then the results are 73.131, 75.797, 77.575 and 79.152, respectively (see Fig. 1). Now let us fix the total number of citations to be 200 for all and consider different values of the total h-index as 3, 7, 12 and 16 then the corresponding A-index for each are, 70.797, 72.547, 73.220 and 73.474, respectively (see Fig. 2).

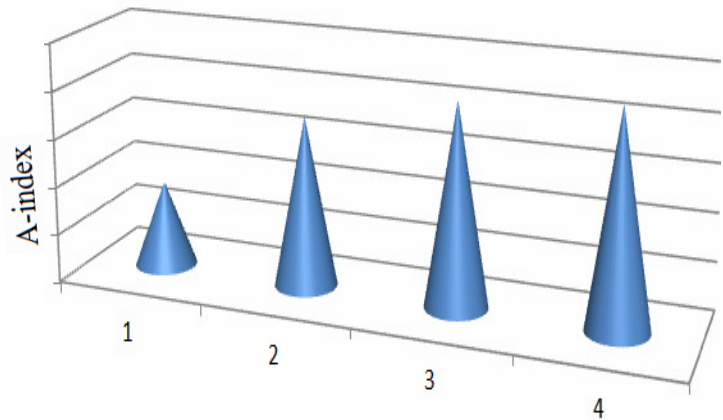

Fig. 2, plot of A-index for different values of the total h-index as 3, 20, 30 and 70.

| Authors | A1 | A2 | A3 | A4 |
|---|---|---|---|---|
| Total $h$-index | 3 | 7 | 12 | 16 |
| $A$-index | 70.797 | 72.547 | 73.220 | 73.474 |

Now we fix all parameters and change the single author h-index as 0, 1, 4, 5 then the corresponding A-index for each author are, 50.631, 65.631, 74.631 and 75.631, respectively (see Fig. 3). These figures show that any small change of the involved factor leads to a dramatically change of the $A$-index. In other words the suggested equation is very sensitive to the Single-author papers $h$-index factor.

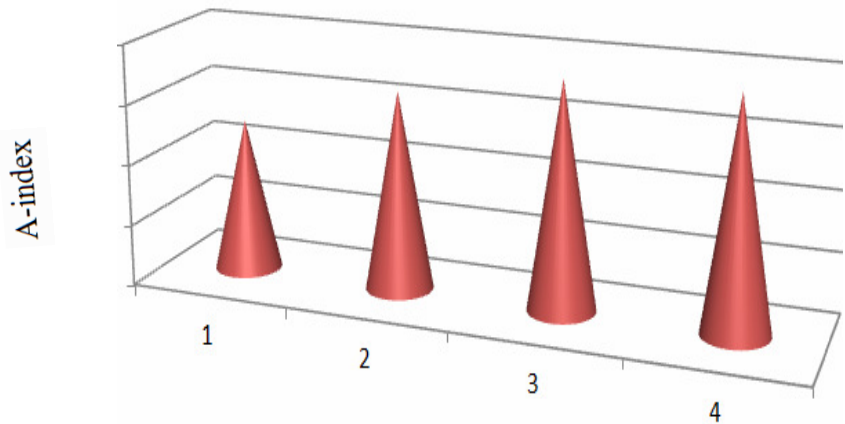

Fig. 3, plot of A-index for different values of the single-author papers h-index as 0, 1, 4 and 5.

| Authors | A1 | A2 | A3 | A4 |
|---|---|---|---|---|
| Single-author papers $h$-index | 0 | 1 | 4 | 5 |
| *A*-index | 50.631 | 65.631 | 74.631 | 75.631 |

From the above results, it is shown that small changes in the single-author papers h-index induceslarge changes in the A-index, while big changes in the number of citations give less change in the A-index. That is, the A-index give quantitative and qualitative indications of the research output and is based on very different factors ----- and gives credit to single-author papers. Inevitably, the diverse perspectives which use citations methods only and their differing views on how people rank different researchers should be studied taking into account disagreement and controversy over how quality should be evaluated.  Despite this, it is seen as important to develop the present formula and to use a common criterion which allows us to evaluate the quality of published work.

One of the biggest challenges may be described as follows: How could we select the main author of the multi-author papers ---to assure the quality and trustworthiness of his/her contribution. More commonly, the researchers' index needs to reflect the quality and justified against the abovementioned criticism and to avoid misleading results.

**Conclusion:** we have studied the efficiency of individual's scientific research output as a function of the single-author's papers-impact and multi-author's papers citations and the corresponding *h*-index, as well as addressed the effect of collective citations and different factors. When any of the variables is increased the suggested index is rising. Using a simple analysis, we have presented comparison between different cases. In particular, we noticed extremely fast convergence for the authors published most of their work independently as single-author papers. This may provide useful tool for quantitatively characterize main difference between the authors who are working jointly and authors who are working indecently. In general, obtaining a deeper understanding of the behavior of the individual's scientific research output needs more discussions and examinations.